\documentclass[11pt,a4paper]{article}
\pdfoutput=1                   
\usepackage{jheppub}
\usepackage{mathtools}         
\usepackage{graphicx}
\graphicspath{{./Figs/}}
\usepackage{comment}
\usepackage{latexsym}
\usepackage{xspace}
\usepackage{lscape}
\usepackage{color}
\usepackage{hyperref}
\usepackage{bm}
\usepackage{bbm}
\usepackage{relsize}
\usepackage{tabularx}
\usepackage{multirow}
\usepackage{amsmath}
\usepackage{amssymb}
\usepackage[table]{xcolor}
\usepackage{lipsum}
\usepackage{feynmp-auto}
\usepackage{upgreek}
\usepackage{nicefrac}
\usepackage[titletoc]{appendix}

\usepackage{caption}
\usepackage{subcaption}

\usepackage{cleveref} 
    \crefname{equation}{}{}
    \crefname{figure}{}{}    
    \crefname{table}{}{}
    \crefname{section}{}{}   
    \crefname{appendix}{}{}
    \crefname{footnote}{}{}
    
    \crefalias{subequation}{equation}

\newcommand{\idx}[1]{{\scriptscriptstyle{(#1)}}}

\newcommand{\lRidx}[1]{{\lambda^{\mkern-3mu\idx{#1}}_{\rm R}}}
\newcommand{\sfrac}[2]{{\textstyle{\frac{#1}{#2}}}} 
\newcommand{\overbar}[1]{\mkern 1.5mu\overline{\mkern-1.5mu#1\mkern-1.5mu}\mkern 1.5mu} 

\def\beq{\begin{equation}}
\def\eeq{\end{equation}}
\def\baq{\begin{eqnarray}}
\def\eaq{\end{eqnarray}}

\DeclareMathOperator{\dalembert}{\square}

\title{Tachyonic production of dark relics:\\
a non-perturbative quantum study}

\author[a,b]{Kimmo Kainulainen}
\author[a,b]{Olli Koskivaara}
\author[a,b]{Sami Nurmi}
\affiliation[a]{Department of Physics, PL 35 (YFL), 40014 University of Jyv\"askyl\"a, Finland}
\affiliation[b]{Helsinki Institute of Physics, PL 64, 00014 University of Helsinki, Finland}
\emailAdd{kimmo.kainulainen@jyu.fi}
\emailAdd{olli.a.koskivaara@student.jyu.fi}
\emailAdd{sami.t.nurmi@jyu.fi}

\abstract{We study production of dark relics during reheating after the end of inflation in a system consisting of a non-minimally coupled spectator scalar field and the inflaton. We derive a set of renormalized quantum transport equations for the one-point function and the two-point function of the spectator field and solve them numerically. We find that our system can embody both tachyonic and parametric instabilities. The former is an expected result due to the non-minimal coupling, but the latter displays new features driven by a novel interplay of the two-point function with the Ricci scalar. We find that when the parametric instability driven by the two-point function takes place, it dominates the total particle production. The quantitative results are also found to be highly sensitive to the model parameters.}


\begin{document}

\maketitle

%
\section{Introduction}
\label{sec:intro}
%

Classical scalar fields coupled to out-of-equilibrium quantum matter play an important role in various settings in cosmology. Some key examples include non-perturbative particle production processes during the reheating after inflation, via a parametric
resonance~\cite{Kofman:1994rk,kofman:1997yn,Greene:1997fu,Braden:2010wd,Berges:2002cz} or via spinodal instability~\cite{Calzetta:1989bj,Guth:1985ya,Weinberg:1987vp,Dufaux:2006ee,Felder:2000hj,Felder:2001kt,Bassett:1997az,Markkanen:2015xuw,Fairbairn:2018bsw}, as well as the processes leading to the electroweak  baryogenesis~\cite{Cline:2000nw,Kainulainen:2001cn,Kainulainen:2002th,Cline:2013gha,Cline:2017qpe,Cline:2020jre,Konstandin:2013caa,Kainulainen:2021oqs} and the leptogenesis mechanism~\cite{Buchmuller:2000nd,Beneke:2010dz,Anisimov:2010dk,Dev:2017trv,DeSimone:2007gkc,Garny:2009qn,Garbrecht:2011aw,Garny:2011hg,Dev:2017wwc,Jukkala:2021sku}. Finding a complete solution of such problems often requires non-perturbative methods and non-equilibrium quantum field theory. In particular in the resonant particle production case the newly created quanta may significantly affect the evolution of the system~\cite{Boyanovsky:1992vi,Boyanovsky:1993pf,Baacke:2001zt,Arrizabalaga:2004iw,Arrizabalaga:2005tf,Kainulainen:2021eki}.

In this work we study tachyonic dark matter production during the reheating epoch in a setup proposed in~\cite{Markkanen:2015xuw,Fairbairn:2018bsw}. Non-minimally coupled scalar fields may undergo a tachyonic instability, or spinodal decomposition, when an effective mass term $\xi R \chi^2$ periodically takes negative values, driven by the oscillating Ricci scalar $R$ during reheating. In~\cite{Markkanen:2015xuw} it was shown that for stable scalar fields with sufficiently weak couplings to visible matter the tachyonic particle production induced by the curvature coupling produces adiabatic dark matter, whose abundance can be made to agree with the observed value over a wide range of coupling values. The results of~\cite{Markkanen:2015xuw} and later in~\cite{Fairbairn:2018bsw} are based on perturbative studies of the particle production similar to those applied to the so called tachyonic reheating in~\cite{Dufaux:2006ee,Felder:2000hj,Felder:2001kt}.~In~\cite{Figueroa:2021iwm}, the dynamics of non-minimally coupled scalars were studied using classical lattice simulations but most of the numerical results shown in that work apply to the case without scalar self-interactions. Here we revisit the tachyonic dark matter production of~\cite{Fairbairn:2018bsw} applying a fully non-perturbative 2PI-approach using methods introduced in~\cite{Kainulainen:2021eki} (for earlier work see~\cite{Herranen:2008hi,Herranen:2008hu,Herranen:2008di,Herranen:2010mh,Fidler:2011yq}).

The 2PI-framework is a powerful tool for studying dynamical non-equilibrium problems. It results in 
evolution equations which naturally include the backreaction from out-of-equilibrium modes on the evolution of the one-point function. We derive the renormalized 2PI equations of motion in an on-shell scheme in terms of physical parameters in the lowest non-trivial loop approximation. We then solve for the coupled dynamics of the one- and two-point functions of the scalar field and investigate the momentum structure of the two-point function. We identify the non-perturbative processes of parametric resonance and spinodal instability taking place during the reheating stage. The efficiency of these processes is found to sensitively depend on the parameters of the theory, such as the spectator self-interaction strength and the inflaton decay rate. Also, the tachyonic and subsequent parametric processes may be coupled in a very intricate way. We note that the methods and their numerical implementation discussed here are not limited to the particular example at hand, but similar techniques can be carried also to more general setups.  

This paper is organized as follows. In section~\cref{sec:model} we introduce the model and in section~\cref{sec:2PI} we derive the renormalized 2PI equations of motion in the comoving frame in the Hartree approximation. In section~\cref{sec:moments} we recast the equation for the two-point function into a form of moment equations in the mixed representation. In section~\cref{sec:results} we apply the numerical approach introduced in~\cite{Kainulainen:2021eki} to the physical setup of~\cite{Fairbairn:2018bsw}, which included backreaction but assumed adiabatic expansion for the mode functions and some further technical approximations. Finally, section~\cref{sec:conclusions} contains our conclusions and outlook.

%
\section{The model}
\label{sec:model}
%

Following~\cite{Markkanen:2015xuw,Fairbairn:2018bsw}, we study a $\mathbb{Z}_{2}$-symmetric scalar singlet model where the singlet $\chi$ has no couplings to other matter fields. The singlet action is given by 
\begin{equation}
\mathcal{S}_{\chi} = \int \mathrm{d}^4x\,\sqrt{-g}\biggl[
          \frac{1}{2}(\nabla^{\mu}\chi)(\nabla_{\mu}\chi) - \frac{1}{2} m^2\chi^2
         + \frac{\xi}{2}R\chi^2-\frac{\lambda}{4}\chi^4 \biggr].
\label{chiaction}
\end{equation}
We use the particle physics convention for the metric signature: $\mathrm{d}s^2 = \mathrm{d}t^2 - a^2\mathrm{d}{\bm{x}}^2$. We will assume that the singlet is energetically subdominant during inflation and reheating, $\rho_{\chi} \ll 3H^2 M_{\rm P}^2$, and treat it as a test field in a classical background space-time, whose evolution is determined by the inflaton field $\phi$. It should be noted that the non-minimal coupling $\xi R\chi^2$ of the field $\chi$ quantized in a classical curved space-time acquires radiative corrections already at the one loop level in the presence of the self-interaction \cite{Buchbinder:1992rb}. Therefore, although $\xi$ can be renormalized to zero at any given scale, it cannot be made to vanish on all scales.

Rescaling the field $\chi$ by the scale factor,
\beq
\sigma \equiv a(t) \chi ,
\eeq
and switching to the conformal time \(\eta\) defined through $a \mathrm{d}\eta \equiv \mathrm{d} t$, we can recast the action \cref{chiaction} for the $\chi$-field into an effectively flat space form:
\beq
\label{eq:sigma_action}
\mathcal{S}_{\sigma}=\int \mathrm{d}\eta \, \mathrm{d}^3\bm{x}\left[\frac{1}{2}(\partial_{\eta}\sigma)^2-\frac{1}{2}(\nabla\sigma)^2
 -\frac{1}{2} m_{\rm eff}^2(\eta) \sigma^2-\frac{\lambda}{4}\sigma^4\right],
\eeq
where the time-dependent effective mass term is defined as
\begin{equation}
m^2_{\rm eff}(\eta) \equiv a^2(\eta)\bigg[m^2 - \bigg( \xi-\frac{1}{6} \bigg)R(\eta)\bigg].
\label{eq:effective_mass}
\end{equation}
This action is the starting point for our derivation of the coupled evolution equations for the one- and two-point functions of the $\chi$-field.

\paragraph{Equation of motion for the inflaton and the scale factor.}

We will treat the inflaton at the classical level and assume a quadratic inflaton potential. Because we wish to study the $\chi$-evolution beyond the decay of the inflaton, we also add a coupling between the inflaton and a radiation component. The radiation energy density is set to zero before the end of inflation. Moreover, we will treat $\chi$ as a test field, so that the Hubble rate and the evolution of the Ricci scalar are determined solely by the inflaton and the radiation component. We then have
\begin{equation}
\begin{aligned}
\ddot{\phi} + 3 H\dot{\phi}+\Gamma \dot\phi+m_{\phi}^2\phi & = 0 ,\\
\dot \rho_{\rm rad} + 4H \rho_{\rm rad} &=  \Gamma \dot \phi^2 ,
\label{eq:inflaton_equations}
\end{aligned}
\end{equation}
where the dots denote differentiation with respect to the cosmic time $t$.

The above equations are solved together with the Friedmann equation $\dot a/a = H$, where the Hubble rate is given by
\begin{equation}
H = \frac{1}{\sqrt{6}M_{\rm P}}\left( \dot\phi^2+    m_\phi^2 \phi^2 + 2\rho_{\rm rad}\right)^{1/2},
\label{eq:friedman} 
\end{equation}
with \(M_{\mathrm{P}}\) being the reduced Planck mass. The time-dependent Ricci scalar in this setup is given by 
\begin{equation}
R = \frac{1}{ M_{\rm P}^2}\left(\dot\phi^2-2 m_\phi^2 \phi^2 \right), 
\label{eq:Ricci}
\end{equation}
as the conformally invariant radiation component gives no contribution at the classical level. These equations can be solved independently of the equations of motion for the spectator field. In the latter the scale factor $a$ and the Ricci scalar $R$ then appear as external functions that source the non-trivial behaviour of the $\chi$-field.

%
\section{The renormalized 2PI equations of motion}
\label{sec:2PI}
%

In this section we derive the renormalized equations of motion for the mean $\sigma$-field and its two-point function corresponding to the action~\cref{eq:sigma_action}, using the 2PI effective action technique of non-equilibrium quantum field theory~\cite{Cornwall:1974vz,Berges:2004yj}. The generic form of the 2PI effective action of a scalar field is
\begin{equation}
\Gamma_{\rm 2PI} [\bar\sigma, \Delta_\sigma] = \mathcal{S}[\bar\sigma] - \frac{\mathrm{i}}{2}\mathrm{Tr}_{\mathcal{C}}\bigl[ \ln (\Delta_\sigma)\bigr] + \frac{\mathrm{i}}{2}\mathrm{Tr}_{\mathcal{C}}\bigl[\Delta_{0\sigma}^{-1}\Delta_\sigma\bigr] 
+ \Gamma_2[\bar\sigma, \Delta_\sigma],
\label{2PI-effective-action}
\end{equation}
where $\mathcal{S}$ is the classical action, $\bar\sigma(x)$ is the classical field and $\Delta_\sigma(x,y)$ is the connected two-point function of the scaled $\sigma$-field and the trace contains integration over the Keldysh contour $\mathcal{C}$~\cite{Keldysh:1964ud} and summation over possible field indices. The classical, real-time inverse propagator is
\begin{equation}
\mathrm{i}\Delta_{0\sigma,ab}^{-1}(x,y;\bar\sigma) = - \Big[ \dalembert_x + m^2_{\rm eff}(\eta) + 3\lambda \bar\sigma_a^2\Big]\delta^{(4)}(x-y)\delta_{ab},
\label{eq:free-inverse-propagator}
\end{equation}
where $\dalembert_x = \partial_\eta^2 -\partial_{\bm x}^2$ and $a,b \in \{1, 2\}$ are the time path indices of the Keldysh contour. The interaction term $\Gamma_2[\bar\sigma,\Delta_\sigma]$ consists of all two-particle irreducible vacuum graphs with lines corresponding to the full propagator $\Delta_\sigma$ and interactions derived from the shifted Lagrangian density $\mathcal{L}[\sigma \rightarrow \bar\sigma + \sigma_q]$, where $\sigma_q$ is the quantum fluctuation around the classical field configuration $\bar \sigma$.

The equations of motion of the one- and two-point functions are then obtained as the stationary conditions of the 2PI effective action:
\begin{equation}
\frac{\delta \Gamma_{\rm 2PI}}{\delta \bar\sigma_a}=0
\qquad {\rm and} \qquad
\frac{\delta \Gamma_{\rm 2PI}}{\delta \Delta_\sigma^{ab}}=0.
\label{eq:stationarity}
\end{equation}
We will be restricting our attention to the lowest non-trivial order in the 2PI-expansion, called the Hartree approximation. In this case the interaction term is just
\begin{equation}
  \Gamma_2[\bar\sigma, \Delta_\sigma] \equiv -\frac{3\lambda}{4} \int \mathrm{d}\eta \, \mathrm{d}^3 \bm{x} \, \Delta_\sigma^2(x,x).
\end{equation}
The non-renormalized equations of motion then become
\begin{subequations}
\label{eq:eom-for-12pt-fununren}
\begin{align}
\biggl[ \dalembert_x + m_{\rm eff}^2(\eta) + \lambda\bar\sigma^2(x)
+ 3 \lambda\Delta_\sigma(x,x)\biggr] \bar\sigma(x) &= 0 ,
\label{eq:eom-for-1pt-fun}
\\
\biggl[ \dalembert_x + m_{\rm eff}^2(\eta) + 3\lambda \bar\sigma^2(x) + 3 \lambda\Delta_\sigma(x,x) \biggr]\mathrm{i}\Delta_\sigma^{ab}(x,y) &= a\delta^{ab} \delta^{(4)}(x-y).
\label{eq:eom-for-2pt-fun}
\end{align}
\end{subequations}
In particular the bare local correlation function $\Delta_\sigma(x,x)$ is a divergent quantity and equations~\cref{eq:eom-for-12pt-fununren} clearly need to be renormalized. We shall now show how this can be done in the 2PI-context, generalizing the derivation of~\cite{Kainulainen:2021eki} to a non-static space-time.

%
\subsection{Renormalization}
\label{sec:renormalization}
%

A systematic renormalization in the 2PI-context was developed in~\cite{Berges:2005hc}, but we shall follow an equivalent, more intuitive method introduced in~\cite{Fejos:2007ec} and extended to curved space-time in~\cite{Arai:2012sh} (see also~\cite{Pilaftsis:2017enx,Pilaftsis:2013xna}). A crucial difference between the 1PI- and the 2PI-cases is that in the latter an infinite number of counterterms and loop diagrams get resummed and mix at high orders in the perturbative expansion. This introduces a number of sub-divergences that may depend on finite temperature or even on the out-of-equilibrium quantum corrections and gives rise to auxiliary $n$-point functions, where some or all of the external field lines are replaced by internal propagators. Each auxiliary function needs a new renormalization condition, but the final equations of motion are independent of the particular choices. We shall closely follow the treatment of~\cite{Kainulainen:2021eki}, extending it to the case of non-zero curvature.

The renormalized quantities are defined from the bare ones through
\begin{equation}
\begin{alignedat}{2}
  \sigma &\equiv Z^{1/2}_\idx{2} \sigma_{\rm R}, \hspace{5em}  
  \Delta_\sigma &&\equiv Z_\idx{0} \Delta_{\rm R}, \\
m_\idx{i}^2 &\equiv m^2_{{\rm R}\idx{i}} + \delta m^2_\idx{i}, \hspace{2.85em} 
\lambda_\idx{i} &&\equiv \lRidx{i} + \delta\lambda^\idx{i},      \hspace{3.5em}
\xi_\idx{i} \equiv \xi_\mathrm{R}^\idx{i} + \delta \xi^\idx{i}.
\label{eq:ren-quant}
\end{alignedat}
\end{equation}
The index enclosed in parenthesis tells how many lines in the vertex function corresponding to the coupling or mass parameter in question are associated with external fields, as explained in~\cite{Fejos:2007ec,Kainulainen:2021eki}. Note that both the bare and the renormalized couplings in general are different for different $i$, as we shall see below. We then define accordingly:
\begin{subequations}
\label{eq:effcts}
\begin{align}
\delta_\lambda^\idx{0} &\equiv Z^2_\idx{0} \big(\lRidx{0} + \delta \lambda^\idx{0}\big) - \lRidx{0},\\
\delta_\lambda^\idx{2} &\equiv Z_\idx{0}Z_\idx{2} \big(\lRidx{2}+ \delta \lambda^\idx{2}\big) - \lRidx{2},\\
\delta_\lambda^\idx{4} &\equiv Z^2_\idx{2} \big(\lRidx{4}+ \delta \lambda^\idx{4}\big) - \lRidx{4},\\
\delta_m^\idx{i} &\equiv Z_\idx{i} \bigl(m^2_{{\rm R}\idx{i}} + \delta m^2_\idx{i}\bigr) - m^2_{{\rm R}\idx{i}},\\
\delta_{\xi}^\idx{i} &\equiv Z_\idx{i} \bigl(\xi_{\mathrm{R}}^{\idx{i}} - \sfrac{1}{6} + \delta \xi^{\idx{i}}  \bigr) - \xi_{\mathrm{R}}^{\idx{i}} + \sfrac{1}{6} .
\end{align}
\end{subequations}
Given these definitions we can write the unrenormalized equations of motion in terms of the renormalized quantities as follows: 
\begin{subequations}
\label{eq:eom-for-12pt-fun}
\begin{align}
\begin{split}
\biggl[ Z_\idx{2} \dalembert_x &+ a^2\Bigl(m^2_{{\rm R}\idx{2}} + \delta_m^\idx{2}\Bigr) 
- a^2 \Big(\xi_\mathrm{R}^\idx{2} - \sfrac{1}{6} + \delta_{\xi}^\idx{2} \Bigr)R \\
&+ 3\Bigl(\lRidx{4} + \sfrac{1}{3}\delta_{\lambda}^\idx{4}\Bigr)\sigma_{\mathrm{R}}^2 
+ 3\Bigl(\lRidx{2} + \delta_{\lambda}^\idx{2}\Bigr)\Delta_{\mathrm{R}}(x,x)\biggr] \sigma_{\mathrm{R}}(x) = 2\lambda_{\rm R}^\idx{4}\sigma^3_{\rm R} \, , 
\end{split}
\label{eq:eom-for-1pt-funR}
\\
\begin{split}
\biggl[ Z_\idx{0} \dalembert_x &+ a^2\Bigl(m^2_{{\rm R}\idx{0}} + \delta_m^\idx{0} \Bigr) 
- a^2 \Big( \xi_{\mathrm{R}}^{{\idx{0}}} - \sfrac{1}{6} + \delta_\xi^\idx{0} \Big)R \\
&+ 3\Bigl(\lRidx{2} + \delta_{\lambda}^\idx{2}\Bigr)\sigma_{\mathrm{R}}^2 
 + 3\Bigl(\lRidx{0} + \delta_{\lambda}^\idx{0}
 \Bigr)\Delta_{\mathrm{R}}(x,x)\biggr]\mathrm{i}\Delta_{\mathrm{R}}^{bc}(x,y) =b\delta^{bc} \delta^{(4)}(x-y).
 \end{split}
\label{eq:eom-for-2pt-funR}
\end{align}
\end{subequations}
Here and in what follows we drop the bar when referring to the classical field $\sigma_{\rm R}$.

\paragraph{Renormalization conditions.}

To proceed, we must now define the renormalization conditions. We start by setting on-shell conditions for the auxiliary two-point function $\Delta^{11}_{\rm R}$ at a vanishing external vacuum expectation value, $\sigma_{\mathrm{R}} = v_{\mathrm{R}} = 0$, and some finite $R = R_0$, along with the requirement that the quantum corrections vanish at the minimum of the effective action:
\begin{equation}
\mathrm{i}\bigl(\Delta^{11}_{\rm R}\bigr)^{{-1}}\bigg|_{\stackrel{\scriptstyle\sigma_{\mathrm{R}}=0}{R=R_0}} \equiv k^2 - a^2m_{\mathrm{\Delta}}^2,  
\quad  
\frac{\rm d}{{\rm d}k^2}\,\mathrm{i}\bigl(\Delta^{11}_{\rm R}\bigr)^{-1}\bigg|_{\stackrel{\scriptstyle\sigma_{\mathrm{R}}=0}{R=R_0}}  \equiv 1
\quad {\rm and} \quad 
\frac{\delta\Gamma_{\rm 2PI}}{\delta \sigma_{\mathrm{R}}}\bigg|_{\stackrel{\scriptstyle\sigma_{\mathrm{R}}=0}{R=R_0}} \equiv 0.
\label{eq:intermediate-ren-conditions}
\end{equation}
Note that we are using the comoving units, so $k$ is also the comoving 4-momentum. These conditions imply that $Z_\idx{0} = 1$. Furthermore, one finds $Z_\idx{2} = 1$ in the Hartree approximation, when the renormalization is performed at $\sigma_{\mathrm{R}} = 0$~\cite{Kainulainen:2021eki}. As a result, one can set also $\smash{m^2_{\mathrm{\Delta}} = m^2_{\rm ph}}$, where $m_{\rm ph}$ refers to the usual mass parameter defined at the off-shell momentum $p^2=0$. The renormalization conditions~\cref{eq:intermediate-ren-conditions}, together with the equation of motion~\cref{eq:eom-for-2pt-funR}, then give
\begin{equation}
m^2_{{\rm R}\idx{0}} + \delta_m^\idx{0} 
- \Big(\xi_\mathrm{R}^\idx{0} - \sfrac{1}{6} + \delta_{\xi}^\idx{0} \Big)  R_0
+ 3\Bigl(\lRidx{0} + \delta_{\lambda}^\idx{0}\Bigr)a^{-2}\Delta_\mathrm{R} = m^2_{\mathrm{ph}}.
\label{eq:eom-GR}
\end{equation}
Here $\Delta_\mathrm{R}$ is computed at the renormalization point. The $a^{-2}$-factor multiplying $\Delta_{\rm R}$ arises from the scaling of the field $\sigma$. In physical units it is absorbed to the correlation function.

In the Hartree approximation we can renormalize $\lambda_{\rm R}^\idx{0}$ and $\lambda_{\rm R}^\idx{2}$ similarly, by setting 
\begin{equation}
\delta_\lambda^\idx{0} \equiv \delta_\lambda^\idx{2}.
\label{eq:rencons-1}
\end{equation}
From $Z_\idx{0,2}= 1$ it then follows that $\lambda_{\rm R}^\idx{0} = \lambda_{\rm R}^\idx{2}$. So, both bare and renormalized couplings can be chosen equal for these vertex functions. Next we set the bare mass parameters $m_\idx{i}^2$ and the $\xi_\idx{i}$-parameters equal for $i \in \{0,2\}$, which gives
\begin{equation}
m^2_{{\rm R}\idx{0}} + \delta_m^\idx{0} = m^2_{{\rm R}\idx{2}} + \delta_m^\idx{2}
\qquad {\rm and} \qquad
\xi_{\rm R}^\idx{0}+\delta_\xi^\idx{0} = \xi_{\rm R}^\idx{2} + \delta_\xi^\idx{2},
\label{eq:rencons-2}
\end{equation}
and we finally define
\begin{equation}
\lRidx{4} + \sfrac{1}{3}\delta_\lambda^\idx{4} \equiv \lRidx{0} + \delta_\lambda^\idx{0}.
\label{eq:rencons-3} 
\end{equation}
This condition ensures that renormalized effective potential has the same first derivative as the tree level potential for a finite $\sigma_{\mathrm R}$ (for more details, see~\cite{Kainulainen:2021eki}). Note that the bare coupling $\lambda_\idx{4}$ is then different from $\lambda_{\idx{0,2}}$, but this has no consequence for the renormalized low-energy theory. Finally, we could relate $\xi_{\mathrm{R}}^\idx{0}$ to a physical mass measured in a background with a non-zero $R$, but we simply define it as an $\overline{\rm MS}$-parameter instead.

\paragraph{Cancellation of the sub-divergences.}

Next we impose the conditions on the cancellation of the sub-divergences~\cite{Fejos:2007ec}. To this end we must work out the primitive divergence in the local correlation function, which in the Hartree approximation is given just by the momentum integral over the renormalized correlator $\mathrm{i}\Delta^{11}_{\rm R}$ defined in the conditions~\cref{eq:intermediate-ren-conditions}:
\begin{align}
\Delta_{\mathrm{R}} &= Q^\epsilon \int \frac{{\rm d}^dp}{(2\uppi )^d}\,\Delta_{\rm R}^{11} (p) 
= -\frac{a^2m_{\mathrm{ph}}^2}{16\uppi^2}\biggl[\frac{2}{\overline{\epsilon}} + 1 - \ln\biggl(\frac{a^2m^2_{\mathrm{ph}}}{Q^2}\biggr)\biggr]
\nonumber \\
&\equiv a^2m_{\mathrm{ph}}^2 \Delta_{\overline\epsilon} + \Delta_{\rm F0}\bigl(am_{\rm ph},Q\bigr),
\label{eq:division-R}
\end{align}
where $\Delta_{\overline\epsilon} \equiv -1/\bigl(8\uppi^2\overline\epsilon\bigr)$ and $Q$ is the comoving momentum scale used for the $\overline{\rm MS}$-re\-nor\-mal\-i\-za\-tion.
Substituting this expression back into equation~\cref{eq:eom-GR} and requiring that the finite and divergent parts cancel separately, we find the following two equations:
\begin{align}
m_{\rm ph}^2 &\equiv m_{{\rm R}\idx{0}}^2 - \Bigl(\xi_\mathrm{R}^\idx{0}-\sfrac{1}{6}\Bigr) R_0 + 3\lRidx{0}a^{-2}\Delta_{\rm F0},
\label{eq:tree-level-def}
\\[.2em]
0 &= \delta_m^\idx{0} - R_0\delta_{\xi}^\idx{0} 
+ 3\delta_\lambda^\idx{0}a^{-2}\Delta_{\rm F0}
+ 3\Bigl(\lRidx{0}+\delta_\lambda^\idx{0}\Bigr)m_{\rm ph}^2\Delta_{\overline\epsilon}. 
\label{eq:loop-level-eq}
\end{align}
Using equation~\cref{eq:tree-level-def} one can rewrite equation~\cref{eq:loop-level-eq} as
\begin{equation}
\begin{split}
\delta_m^\idx{0} 
 + 3m^2_{{\rm R}\idx{0}}\Big(\lRidx{0}+\delta_\lambda^\idx{0}\Big)\Delta_{\overline\epsilon} 
&+ 3\Big[\delta_\lambda^\idx{0} + 3\Big(\lRidx{0}
       +\delta_\lambda^\idx{0}\Big)\lRidx{0}\Delta_{\overline\epsilon} \Big] 
       a^{-2}\Delta_{\rm F0}
\\  
&- \Big[ \delta_{\xi}^\idx{0} + 3\Bigl(\xi_\mathrm{R}^\idx{0}-\sfrac{1}{6}\Bigr)            
         \Big(\lRidx{0}+\delta_\lambda^\idx{0}\Big)\Delta_{\overline\epsilon} \Big] R_0 = 0.
\label{eq:loop-level-rearr}
\end{split}
\end{equation}
This equation can hold for arbitrary $R_0$ and $\Delta_{\rm F0}$ only if the coefficients multiplying each of these terms vanish separately. This gives us three constraints between the counterterms:
\begin{subequations}
\label{eq:last-ren-eqs2}
\begin{align}
\delta_m^\idx{0} + 3m^2_{{\rm R}\idx{0}}\Bigl(\lRidx{0}+\delta_\lambda^\idx{0}\Bigr)
                                        \Delta_{\overline\epsilon} &= 0,
\label{eq:last-ren-eqs-12} 
\\[.2em] 
\delta_\lambda^\idx{0} + 3\Bigl(\lRidx{0} +\delta_\lambda^\idx{0}\Bigr)\lRidx{0}
                    \Delta_{\overline\epsilon} &= 0,
\\[.2em] 
\delta_{\xi}^\idx{0} + 3\Bigl(\xi_{\mathrm{R}}^\idx{0}-\sfrac{1}{6}\Bigr) \Bigl(\lRidx{0}+\delta_\lambda^\idx{0}\Bigr)\Delta_{\overline\epsilon} &= 0.
\label{eq:last-ren-eqs-22}
\end{align}
\end{subequations}
From these we find the explicit expressions for the counterterms $\delta_\lambda^\idx{0}$, $\delta_m^\idx{0}$ and $\delta_{\xi}^\idx{0}$:
\begin{equation}
\delta_\lambda^\idx{0} 
= - \frac{3\bigl(\lRidx{0}\bigr)^2\Delta_{\overline\epsilon}}{1+3\lRidx{0}\Delta_{\overline\epsilon}},
\hspace{1.5em}
\delta_m^\idx{0} = 
-\frac{3m^2_{{\rm R}\idx{0}}\lRidx{0}\Delta_{\overline\epsilon}}{1+3\lRidx{0}
\Delta_{\overline\epsilon}},\hspace{1.5em} 
\delta_\xi^\idx{0} =  -\frac{3\bigl(\xi_{\mathrm{R}}^\idx{0}-\sfrac{1}{6}\bigr)\lRidx{0}
                  \Delta_{\overline\epsilon}}{1+3\lRidx{0}\Delta_{\overline\epsilon}} \, .
\label{eq:auxiliary-cts}
\end{equation}
The running of the renormalized parameters now follows from requiring that the corresponding bare parameters are constants: $\partial_Q\bigl[Q^\epsilon\bigl(\lRidx{0} + \delta_\lambda^\idx{0}\bigr)\bigr] = 0$, $\partial_Q\bigl[Q^\epsilon\bigl( m^2_{{\rm R}\idx{0}} + \delta_m^\idx{0}\bigr)\bigr] = 0$ and $\smash{\partial_Q\bigl[Q^\epsilon\bigl( \xi^{\idx{0}}_{{\rm R}}\! - \sfrac{1}{6} +\delta_\xi^\idx{0}\bigr)\bigr] = 0}$. For the running of $\lRidx{0}$ and $\xi_{\mathrm R}^\idx{0}$ one then finds
\begin{equation}
\lRidx{0}(Q) 
= \frac{\lambda^\idx{0}_{{\rm R0}}}
                    {1+\frac{3\lambda^\idx{0}_{{\rm R0}}}{8\uppi^2}
                       \ln\Bigl(\frac{Q_0}{Q}\Bigr)}  
\qquad {\rm and} \qquad 
\xi_{\rm R}^\idx{0}(Q) - \sfrac{1}{6} = \frac{\xi_{\rm R0}^\idx{0}-\sfrac{1}{6}}
                                       {1+\frac{3\lambda^\idx{0}_{{\rm R0}}}{8\uppi^2}
                                          \ln\Bigl(\frac{Q_0}{Q}\Bigr)},
\label{eq:running-parameters}
\end{equation}
where $\lambda^\idx{0}_{{\rm R0}}\equiv\lambda^\idx{0}_{{\rm R}}(Q_0)$ and $\xi^\idx{0}_{{\rm R0}}\equiv \xi^\idx{0}_{{\rm R}}(Q_0)$ and our previous choices imply that $\lRidx{2}=\lRidx{0}$. The running of the mass terms is analogous to the running of the couplings~\cite{Kainulainen:2021eki}. On the other hand, the coupling $\lRidx{4}$ does not run at all. Indeed, $\lRidx{4}$ remains finite because of the condition $\delta_\lambda^\idx{4} = 3\delta_\lambda^\idx{0}$ up to finite terms, which implies that $\partial_Q\lRidx{4} = 0$.

\paragraph{Renormalized equations of motion.}

Next we show that the full evolution equations~\cref{eq:eom-for-12pt-fun} get renormalized by the counterterms we have defined. We begin by defining a finite effective mass term, which includes general corrections from $R$, $\sigma_{\rm R}$ and $\Delta_{\rm F}$, as follows: 
\begin{equation}
M_{\rm eff}^2(\sigma_{\mathrm{R}},\Delta_{\rm F}) \equiv a^2\Big[m^2_{{\rm R}\idx{0}} 
- \Big(\xi_{\mathrm{R}}^\idx{0}-\sfrac{1}{6}\Big)R\Big] 
+ 3\lRidx{0}\Bigl(\sigma_{\rm R}^2 + \Delta_{\rm F}\Bigr).
\label{eq:effective-mass-full}
\end{equation}
The finite part $\Delta_{\rm F}$ of the local correlation function $\Delta_{\rm R}$ is defined similarly to 
equation~\cref{eq:division-R}:
\begin{equation}
\Delta_{\mathrm{R}} \equiv M_{\rm eff}^2(\sigma_{\mathrm{R}},\Delta_{\rm F}) 
\Delta_{\overline\epsilon} + \Delta_{\rm F}.
\label{eq:division-full}
\end{equation}
We furthermore split $\Delta_{\rm F} \equiv \Delta_{\rm F0}(M_{\rm eff},Q) + \delta \Delta_{\rm F}$, where $\Delta_{\rm F0}$ was defined in equation~\cref{eq:division-R} and $\delta \Delta_{\rm F}$ represents the remaining non-equilibrium fluctuations. Using this expression, the equation of motion for the two-point function becomes 
\begin{equation}
\begin{split}
\phantom{H}
\biggl[\dalembert_x + M_{\rm eff}^2
&+ a^2\left(\delta_m^\idx{0} - R\delta_{\xi}^\idx{0}\right) 
+ 3\delta_\lambda^\idx{0}\Bigl( \sigma_{\rm R}^2 + \Delta_{\rm F}\Bigr) 
 \\
&+ 3\Bigl(\lRidx{0}+\delta_\lambda^\idx{0}\Bigr)M_{\rm eff}^2
\Delta_{\overline\epsilon}\biggr]\mathrm{i}\Delta^{bc}_{\rm R}(x,y) 
= b\delta^{bc}\delta^{(4)}(x-y).
\label{eq:loop-level-eq-full}
\end{split}
\end{equation}
Using the definition~\cref{eq:effective-mass-full} again in the term proportional to $\Delta_{\overline\epsilon}$, we can write equation~\cref{eq:loop-level-eq-full} as
\begin{equation}
\begin{split}
\phantom{H}
\biggl\{\dalembert_x + M_{\rm eff}^2
&-a^2 \left[\delta_{\xi}^\idx{0} + 3\Bigl(\xi_{\mathrm{R}}^\idx{0}-\sfrac{1}{6}\Bigr) \Bigl(\lRidx{0}+\delta_\lambda^\idx{0}\Bigr)\Delta_{\overline\epsilon} \right] R
 \\
&+ 3\Big[ \delta_\lambda^\idx{0} + 3\Bigl(\lRidx{0} +\delta_\lambda^\idx{0}\Bigr)\lRidx{0}                    \Delta_{\overline\epsilon} \Big] \Bigl( \sigma_{\rm R}^2 + \Delta_{\rm F}\Bigr)
 \\
&+ a^2\Big[\delta_m^\idx{0} + 3m^2_{{\rm R}\idx{0}}\Bigl(\lRidx{0}+\delta_\lambda^\idx{0}\Bigr)                                        \Delta_{\overline\epsilon}\Big] \biggr\}\mathrm{i}\Delta^{bc}_{\rm R}(x,y)
= b\delta^{bc}\delta^{(4)}(x-y).
\label{eq:loop-level-eq-full2}
\end{split}
\end{equation}
The renormalization conditions~\cref{eq:last-ren-eqs2} set all the terms in the square brackets to zero leaving behind only the finite mass term $M_{\rm eff}^2$. It should be appreciated how the {\em constant} counterterms cancel infinities that depend on the dynamical variables $\sigma_{\rm R}$, $R$ and $\Delta_{\rm F}$.

Similar manipulations can be done, crucially dependent on the definition~\cref{eq:rencons-3}, in the equation~\cref{eq:eom-for-1pt-funR} for the one-point function. Our final equations then become
\begin{subequations}
\label{eq:ds_eoms}
\begin{align}
\Big[ \dalembert_x + M_{\rm eff}^2(\sigma_{\mathrm{R}},\Delta_{\rm F}) \Big] \sigma_{\mathrm{R}} &= 2\lRidx{4}\sigma_{\rm R}^3,
\label{eq:eom-for-phi-R}
\\
\phantom{H}
\Big[ \dalembert_x + M_{\rm eff}^2(\sigma_{\mathrm{R}},\Delta_{\rm F}) \Big]\mathrm{i}\Delta_{\rm R}^{ab}(x,y) 
&= b\delta^{bc}\delta^{(4)}(x-y).
\label{eq:eom-for-delta-R}
\end{align}
\end{subequations}
Let us finally point out that these equations are independent of the renormalization scale for the auxiliary renormalization conditions: one can show that $\partial_Q (M_{\rm eff}^2)=0$ using the gap equation~\cref{eq:effective-mass-full} together with the running equations~\cref{eq:running-parameters}.

\paragraph{Physical parameters.}

We have now renormalized our equations of motion, but we still have not related our parameters to observable quantities. We now address this problem for completeness, even though none of the parameters in the problem are directly observable. We start by specifying the Hartree-corrected effective potential in the limit of constant curvature, consistent with our renormalization conditions. The calculation is identical to the one given in~\cite{Kainulainen:2021eki} and we only quote the final result, first found in~\cite{AmelinoCamelia:1992nc}:
\begin{equation}
V_{\rm H}(\sigma_{\rm R}) = -\frac{\lRidx{4}}{2}\sigma^4_{\rm R} + \frac{\overbar m^4(\sigma_{\rm R})}{12\lRidx{0}} -\frac{\overbar m^4(\sigma_{\rm R})}{64\uppi^2}\biggl[\ln\biggl(\frac{\overbar m^2(\sigma_{\rm R})}{Q^2}\biggr) - \frac{1}{2} \biggr],
\label{2PI-effective-potential-final}
\end{equation}
where $\overbar m^2$ is the solution to equation~\cref{eq:effective-mass-full} for $R=R_0$ and $\Delta_{\rm F} = \Delta_{\rm F0}\bigl(\overbar m^2\bigr)$. Now, differentiating the effective potential twice, we find
\begin{equation}
\Gamma_{\rm 1PI}^\idx{2}\bigl(p^2 = 0, \sigma_{\mathrm{R}}\bigr) \; = \; \frac{\partial^2 V_{\rm H}(\sigma_{\mathrm{R}})}{\partial \sigma_{\rm R}^2} 
= \overbar m^2(\sigma_{\rm R}) + 6\Bigl[ \lRidx{0}\bigl(\overbar m^2(\sigma_{\rm R})\bigr) - \lRidx{4} \Bigr]\sigma_{\rm R}^2. 
\label{eq:second-derivative-of-potential}
\end{equation}
Because $\smash{\overbar m^2(0) \equiv a^2m_{\mathrm{ph}}^2}$, we see that the mass parameter $m_{\mathrm{ph}}$ of the auxiliary propagator equals the value of the full two-point function $\smash{\Gamma_{\rm 1PI}^\idx{2}\bigl(p^2= 0,\sigma_{\rm R}=0\bigr)}$. Equation~\cref{eq:second-derivative-of-potential} also suggests that it is natural to define $\lRidx{0}(m_\mathrm{\rm ph}) \equiv \lRidx{4}$.

Finally, one can easily show that $\lambda^\idx{4}_{\rm R}$ coincides with the four-point function measured at zero momentum:
\begin{equation}
\lambda_{\mathrm{R}} \; \equiv \;\Gamma_{\rm 1PI}^\idx{4}(p_i=0,\sigma_{\rm R}=0) \;= \; \frac{1}{6}\frac{\partial^4V_{\rm H}(\sigma_{\rm R})}{\partial \sigma_{\rm R}^4}\bigg|_{\sigma_{\rm R}=0}  =  \lRidx{4}.
\label{eq:effective-coupling}
\end{equation}
The mass $m_{\rm ph}$ and the coupling $\lambda_{\rm R}$ can be related to an on-shell mass and a four-point function in the physical region without further reference to the 2PI-methods. Finally, we define the parameter $\xi^{\idx{0}}_{\rm R}$ as the $\overline{\rm MS}$-parameter at scale $m_{\rm ph}$: $\bar \xi_{\rm R} \equiv \xi^{\idx{0}}_{\rm R}(m_{\rm ph})$. These considerations now uniquely define all the parameters in our model.

%
\section{Wigner-space and moment equations}
\label{sec:moments}
%

The direct numerical implementation of equations~\cref{eq:ds_eoms} would be very difficult and we shall use the phase space picture instead. To this end we define the Wigner transform of a generic function of two variables $\mathcal{O}(u,v)$ as follows:
\begin{equation}
\label{eq:wigner}
\mathcal{O}(k,X) \equiv \int\mathrm{d}^4r\,\mathrm{e}^{\mathrm{i}k \cdot r}\,\mathcal{O}\left(X+\frac{r}{2}, X-\frac{r}{2}\right),
\end{equation}
where $r = u-v$ and $X = \frac{1}{2}(u+v)$ are the relative and average coordinates, respectively.
For a homogeneous and isotropic system relevant here, the transformation with respect to spatial coordinates reduces to the ordinary Fourier transformation. In this case the equation~\cref{eq:eom-for-delta-R} for the two-point function in Wigner-space becomes just
\begin{equation}
\label{eq:sd_wigner_des}
\left[\frac{1}{4}\partial_{\eta}^2-k^2-\mathrm{i}k_0 \partial_{\eta} + {{M_{\mathrm{eff}}^2}}\bigl(\eta-\sfrac{\mathrm{i}}{2}\partial_{k_0}\bigr)\right]\mathrm{i}\Delta^{bc}_{\bm k}(k_0,\eta)  = b\delta^{bc},
\end{equation}
where we denoted $M_{\rm eff}^2(\sigma_{\mathrm{R}},\Delta_{\rm F}) \equiv  M_{\mathrm{eff}}^2(\eta)$.

To study the dynamics of the coupled system of the one- and two-point functions it suffices to concentrate on any of the four components of the propagator $\Delta^{ab}$. We choose to work with $\Delta^{+-} = \Delta^<$ and define its $n$\textsuperscript{th} moment as
\begin{equation}
\label{eq:moment}
\rho_{n{\bm k}} \equiv \int\frac{\mathrm{d}k_0}{2\uppi}\,k_0^n\,\Delta^<_{\bm{k}}(k_0,\eta).
\end{equation}
Integrating equation~\cref{eq:sd_wigner_des} over $k_0$, weighted by $1$ and by $k_0$, and taking real and imaginary parts of the resulting equations one finds a closed set of equations for the three lowest moments with $n \in \{0,1,2\}$~\cite{Herranen:2010mh,Kainulainen:2021eki}.~The equation for $\rho_{1{\bm k}}$ is simple: $\partial_\eta\rho_{1{\bm k}}=0$, which implies that $\rho_{1{\bm k}}$ is a constant. In addition we  observe that the quantity
\begin{equation}
\label{eq:X_stab}
X_{\bm k} \equiv 2\rho_{0{\bm k}} \rho_{2{\bm k}}  - \Bigl( |\bm{k}|^2 + M^2_{\mathrm{eff}} \Bigr) \rho_{0{\bm k}}^2  - \sfrac{1}{4}\left(\partial_{\eta} \rho_{0{\bm k}}\right)^2
\end{equation}
is conserved in our setup: $\partial_{\eta} X_{\bm k} = 0$. This is no longer true in an interacting system~\cite{Herranen:2010mh,Kainulainen:2021eki}, but even then using $X_{\bm k}$ as a variable instead of $\rho_{2{\bm k}}$ leads to numerically more stable equations.

In the end we then have the following equations for the homogeneous field $\sigma_{\rm R}$ and the moments $\rho_{n{\bm k}}$:
\begin{equation}
\begin{split}
\Bigl(\partial_{\eta}^2 + {M_{\mathrm{eff}}^2}\Bigr)\sigma_{\rm R} 
&= 2\lambda_{\rm R}\sigma^3_{\rm R},
\\[.2em]
\Bigl(\sfrac{1}{4}\partial_{\eta}^2 + |\bm{k}|^2 + {M_{\mathrm{eff}}^2} \Bigr)\rho_{0{\bm k}} 
&= \rho_{2{\bm k}},
\label{eq:finaleom_desc_stab}
\end{split}
\end{equation}
where $\rho_{2{\bm k}}$ is evaluated using equation~\cref{eq:X_stab}. The non-trivial nature of the evolution equations is hidden in the gap equation~\cref{eq:effective-mass-full}, which couples all the variables. Using the moments and the fact that $M_\mathrm{eff}^2$ is actually $Q$-independent, we can write the gap equation directly in terms of our chosen physical parameters, choosing $Q=am_{\rm ph}$:
\begin{equation}
\begin{split}
M_{\rm eff}^2 
= a^2m^2_{\rm ph} &- a^2\Big(\bar \xi_{\mathrm{R}}-\sfrac{1}{6}\Big)(R-R_0)
                        + 3\lambda_{\rm R}\sigma_{\rm R}^2
+ 3\lambda_{\rm R}\int_{\bm k} \nolimits \Biggl( \rho_{0\bm{k}} - \frac{\Theta_{\bm k}}{2\omega_{{\bm k}}} \Biggr)
\\
&+ \frac{3\lambda_{\rm R}}{16\uppi^2}
 \left[M_{\mathrm{eff}}^2 \ln\left(\frac{M^2_\mathrm{eff}}{a^2m_{\rm ph}^2}\right) 
      - M_{\mathrm{eff}}^2 + a^2m_{\rm ph}^2 \right],
\label{eq:effective-mass-full_wigner}
\end{split}
\end{equation}
where we defined $\int_{\bm k} \equiv \frac{1}{2\uppi^2}\int_0^{\infty} {\rm d}|{\bm k}| |{\bm k}|^2$, $\Theta_{\bm k} \equiv \theta\bigl(\omega_{\bm k}^2(t)\bigr)$, $\omega_{\bm k}^2 \equiv |{\bm k}|^2 + M_{\rm eff}^2$, $\bar\xi_{\mathrm{R}} \equiv \xi_{\rm R}^\idx{0}(m_{\rm ph})$ and $R_0$ is the background Ricci scalar at the renormalization point.\footnote{To get to equation~\cref{eq:effective-mass-full_wigner} one uses for example the relation 
$\smash{m}^2_{\mathrm{R}\idx{0}} = m^2_{\mathrm{ph}}\bigl(1 + \frac{3\lambda_{\mathrm{R}}}{16\uppi^2}\bigr) + \bigl(\bar \xi_{\mathrm{R}}-\sfrac{1}{6}\bigr)R_0$, which can be derived from equation~\cref{eq:tree-level-def} and the running equations for the mass and the couplings.}~We assume that renormalization is performed in a background with no curvature and set $R_0=0$ here.

Finally, we define the particle number density and the quantum coherence functions in terms of the moments as follows~\cite{Herranen:2010mh,Kainulainen:2021eki}:
\begin{subequations}
\label{f-rho_HOM}
\begin{align}
n_{\bm{k}} &\equiv \frac{1}{\omega_{\bm k}}\rho_{2\bm{k}} + \rho_{1\bm{k}},
\label{f-rho_HOM-n}\\
\overbar{n}_{\bm{k}} &\equiv \frac{1}{\omega_{\bm k}}\rho_{2\bm{k}} - \rho_{1\bm{k}} - 1,
\label{f-rho_HOM-nbar}\\
f^{c\pm}_{\bm k} &\equiv \omega_{\bm k}\rho_{0\bm{k}} - \frac{1}{\omega_{\bm k}}\rho_{2\bm{k}} 
\pm \frac{\mathrm{i}}{2}\partial_t\rho_{0\bm{k}}.
\label{f-rho_HOM-fc}
\end{align}
\end{subequations}
We will denote the momentum-integrated versions of these functions by $n \equiv \int_{\bm k} n_{\bm k}$ and $f^c \equiv \int_{\bm k} |f^{c\pm}_{\bm k}|$. In our case of a real field with no collisions $\rho_{1 \bm k} = -1/2$ throughout, so that $n_{\bm{k}}$ and $\overbar{n}_{\bm{k}}$ actually coincide. The functions $f^{c\pm}_{\bm k}$ in turn measure the degree of quantum coherence, or squeezing, between particle-antiparticle pairs with opposite 3-momenta~\cite{Fidler:2011yq}, and particle production can only take place when $f^{c\pm}_{\bm k} \neq 0$. The unique vacuum which corresponds to a state with no particles nor any coherence can then be defined as
\begin{equation}
\rho_{0\bm k}^{\rm vac} \equiv \frac{\Theta_{\bm k}}{2\omega_{\bm k}}, \qquad 
\partial_t\rho_{0\bm k}^{\rm vac}\equiv0, \qquad
\rho_{1\bm k}^{\rm vac} \equiv -\frac{1}{2}
\quad \mathrm{and} \quad 
\rho_{2\bm k}^{\rm vac} \equiv \frac{\omega_{\bm k}}{2}\Theta_{\bm k}.
\label{eq:non-coherent-vacuum}
\end{equation}
The Heaviside theta function $\Theta_{\bm k}$ ensures that no spinodal modes are included in the vacuum. Finally, we define the non-equilibrium fluctuations in the moments as $\delta\rho_{n \bm k} \equiv \rho_{n \bm k} - \rho_{n\bm k}^{\rm vac}$.

%
\section{Results}
\label{sec:results}
%

We numerically solve the equations~\cref{eq:finaleom_desc_stab} and~\cref{eq:effective-mass-full_wigner}, following the methods of~\cite{Kainulainen:2021eki}. We focus on a setup where the energy density of $\sigma$ stays negligible compared to the total energy density, $\rho_{\sigma}\ll 3 H^2 M_{\rm P}^2$, during the entire simulation time. The scale factor $a$ and the Ricci scalar $R$ are therefore entirely set by the inflaton and its decay products via equations~\cref{eq:inflaton_equations}, and they appear as externally given functions in equations~\cref{eq:finaleom_desc_stab,eq:effective-mass-full_wigner}. We choose $m_{\phi} = 1.5 \times 10^{13}$ GeV and set slow roll initial conditions with $\phi_{\rm in} = 15 M_{\rm P}$ on the inflaton sector. On the spectator sector we set $m_{\rm ph} = 150$ GeV, initialize the two-point function $\Delta_{\rm R, in}$ by giving the Minkowski vacuum values~\cref{eq:non-coherent-vacuum} for the moments, and give a small non-zero initial value for the one-point function $\sigma_{\rm R, in}$. In the following,  we denote by $\eta_0$ the moment when $\epsilon_{\rm H}\equiv -\dot{H}/H^2=1$ for the first time.
Our main results are summarized in the figures of this section. 

%
\begin{figure}[t]
     \centering
     \includegraphics[width=0.95\linewidth]{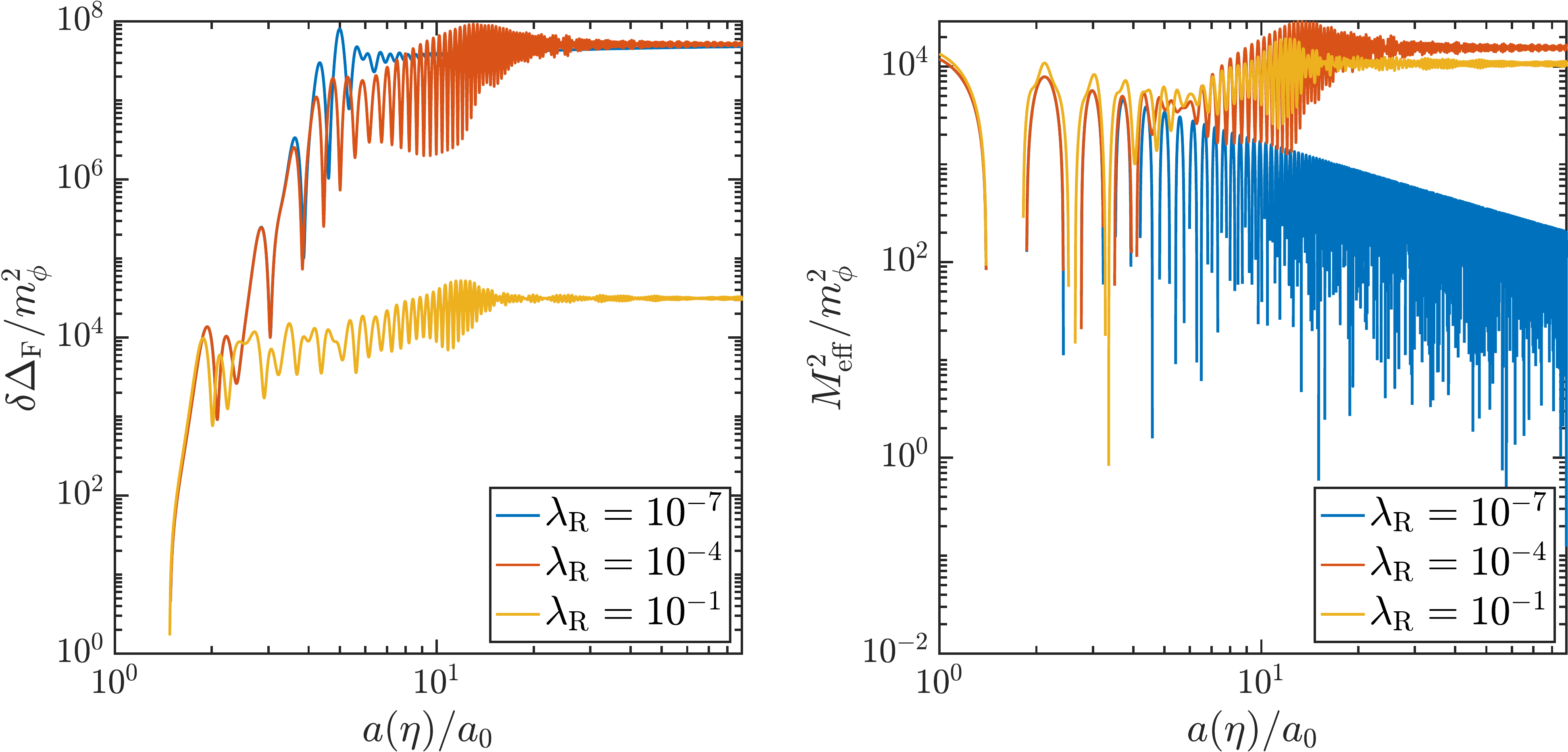} 
        \caption{The two-point function $\delta\Delta_{\mathrm{F}}$  (left panel) and the effective mass function \(M^2_{\mathrm{eff}}\) (right panel). The results are shown for $\lambda_{\mathrm{R}} \in \{10^{-7}, 10^{-4}, 10^{-1}\}$, $\bar\xi_{\mathrm{R}} = 50$ and $\Gamma = 0$.}
        \label{fig:variance1}
\end{figure}
%

\begin{figure}[t]
\centering
\includegraphics[width=0.95\linewidth]{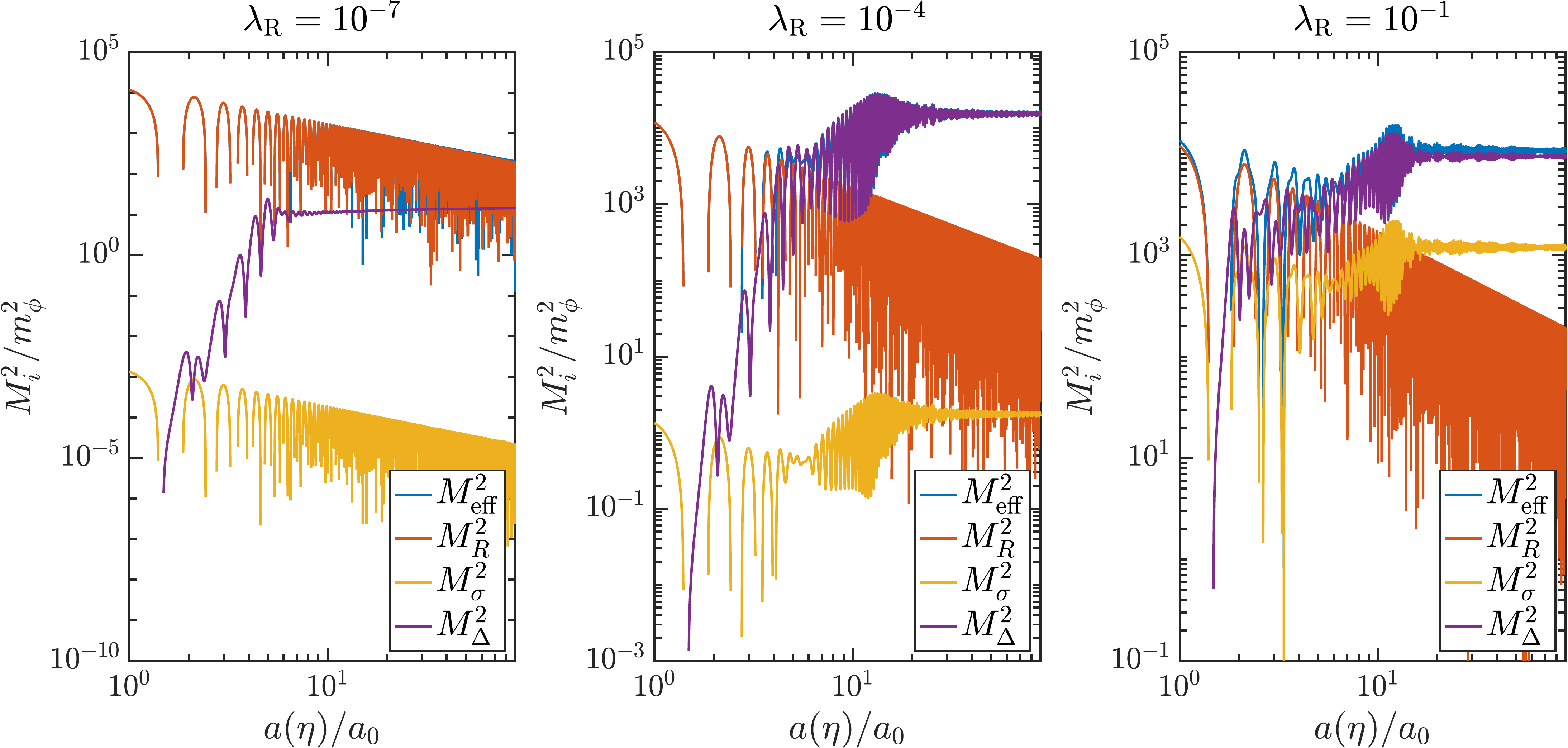}
\caption{The effective mass function $M^2_{\mathrm{eff}}$ (blue) and its component functions $M^2_{R}$ (red), $M^2_{\Delta}$ (violet) and $M^2_{\sigma}$ (yellow), defined in equations~\cref{eq:Meff_components}, for $\lambda_{\mathrm{R}} \in \{10^{-7}, 10^{-4}, 10^{-1}\}$, $\bar\xi_{\mathrm{R}} = 50$ and $\Gamma = 0$.}
\label{fig:meffcomponents}
\end{figure}

\paragraph{Case I: {$\bm{\bar{\xi}_{\mathrm{R}} = 50, \Gamma = 0}$.}}
We will first discuss a case with a non-minimal coupling $\bar{\xi}_{\mathrm{R}} = 50$ and a non-interacting inflaton, $\Gamma = 0$, where the results can be directly compared with those obtained in~\cite{Fairbairn:2018bsw}.~The left panel in figure~\cref{fig:variance1} shows the time evolution of the fluctuation in the contact limit for the comoving two-point function $\langle\sigma_{\rm R}^2\rangle$: $\delta \Delta_{\mathrm F} \equiv \Delta_{\rm F} - \Delta_{\mathrm{F}0}$. The right panel shows the effective mass function $M_{\rm eff}^2$ given by equation \cref{eq:effective-mass-full_wigner}. In both panels the self-coupling is given the values $\lambda_{\rm R} = 10^{-7}$ (blue lines), $10^{-4}$ (red lines) and $10^{-1}$ (orange lines). There are three components of different origin contributing to the effective mass function $M_{\rm eff}^2$:
\begin{subequations}
\label{eq:Meff_components}
\begin{align}
M^2_R        &\equiv -a^2\bigl(\bar{\xi}_{\mathrm{R}}-\sfrac{1}{6}\bigr)R      & {\rm (curvature)},\\[.8em]
M^2_{\Delta} &\equiv  3\lambda_{\mathrm{R}}\delta\Delta_{\rm F} 
                     = 3\lambda_{\mathrm{R}}{ \int_{\bm k}} \delta \rho_{0 \bm k} 
                       & {\rm (fluctuations)}, \\
M^2_{\sigma} &\equiv  M^2_{\rm eff}- M^2_R -  M^2_{\Delta}  & {\rm (field \; and \; background) }.
\end{align}
\end{subequations}
The evolution and magnitudes of these components are displayed in figure~\cref{fig:meffcomponents}. 
%
%
%

For all three values of $\lambda_{\rm R}$ shown in the figures, the field-dependent mass term $M_\sigma^2$ is very small compared to the curvature and fluctuation corrections. In all cases the initial evolution is characterized by a rapid growth of the fluctuation contribution to the two-point function $\delta \Delta_{\mathrm F}$, which is driven by periodic tachyonic instabilities that occur when $M_{\rm eff}^2 < 0 $. The growing two-point function gives a positive definite contribution to the fluctuation part $M^2_\Delta$ in the effective mass function, which is known to eventually terminate the strong tachyonic growth~\cite{Dufaux:2006ee}. 

As seen in figure~\cref{fig:meffcomponents}, for $\lambda_{\rm R} = 10^{-7}$ the growth of $\delta\Delta_{\rm F}$ stops while the effective mass is still dominated by the curvature term, $\langle M^2_\Delta+M^2_\sigma  \rangle_{\rm osc} \!\approx \! \langle M^2_\Delta \rangle_{\rm osc} \!\ll\! \langle M^2_{R} \rangle_{\rm osc}$, where the brackets $\langle \dots \rangle_{\rm osc}$ denote averaging over an oscillation cycle of the mean field $\sigma_{\rm R}$. The reason for this ending of the tachyonic growth is that the windows with $M_{\rm eff}^2 < 0$ become too narrow to generate a coherent net particle production. This effect is controlled by the evolution of $R$, whose oscillation period is a constant in physical time, proportional to the inverse inflaton mass $\smash{m_{\phi}^{-1}}$,
but whose magnitude decreases rapidly, $R\propto a^{-3}$. The time available for tachyonic evolution per oscillation period then shrinks, while the oscillatory evolution between pulses grows, mixing growing and decaying modes. Eventually the tachyonic pulses lose all coherence and no net growth is registered. As a result our final value of $\delta \Delta_{\rm F}$ is about an order of magnitude smaller than in~\cite{Fairbairn:2018bsw}\footnote{Note that our results are expressed in terms of the comoving field $\sigma = a \chi$ while~\cite{Fairbairn:2018bsw} uses the physical field $\chi$. We have normalized the scale factor to $a_0 =12.6$.}, where the tachyonic growth was observed to continue up to $\langle M^2_\Delta \rangle_{\rm osc} \sim \langle M^2_{R} \rangle_{\rm osc}$.~This effect is spurious however, following from the use in~\cite{Fairbairn:2018bsw} of the adiabatic expansion in the regions where the adiabaticity condition $|\dot{\omega}/\omega^2|\ll 1$ for the mode function frequencies no longer holds between the tachyonic windows.

\begin{figure}[t]
\centering
\includegraphics[trim={3.4cm 0 3.4cm 0},clip,width=0.95\linewidth]{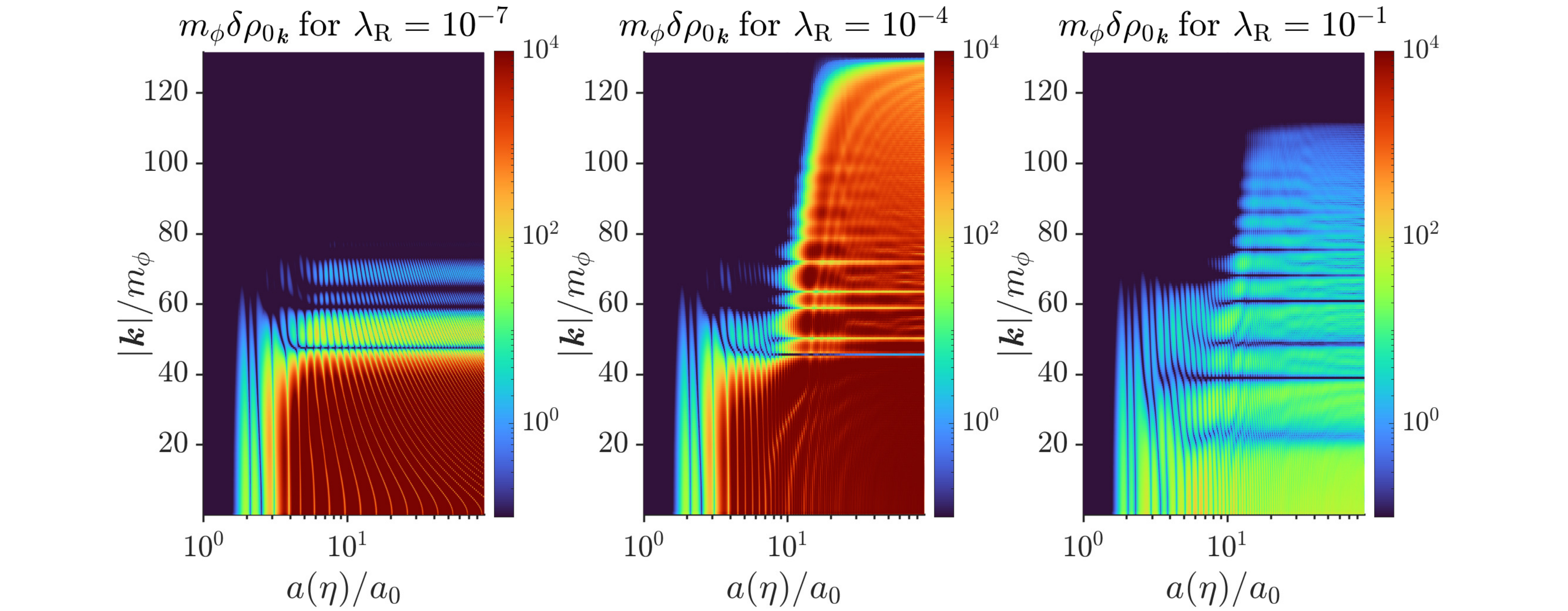}
\caption{The zeroth moment $\delta\rho_{0 \bm k}$ of the two point function  for $\lambda_{\mathrm{R}} \in \{10^{-7}, 10^{-4}, 10^{-1}\}$ with $\bar\xi_{\mathrm{R}} = 50$ and $\Gamma = 0$.}
\label{fig:rho0k}
\end{figure}

The case with larger couplings $\lambda_{\rm R} = 10^{-4}$ and $10^{-1}$ is markedly different. Here the (mostly) tachyonic growth {\em does} continue until $\langle M^2_\Delta \rangle_{\rm osc} \sim \langle M^2_{R} \rangle_{\rm osc}$, after which $\delta \Delta_{\rm F}$ starts to backreact into the dynamics of the system. The evolution of $R$ is exactly the same as in the previous case but the larger coupling $\lambda_{\rm R}$ makes $\langle M^2_\Delta\rangle_{\rm osc}$ bigger, and the backreaction limit $\langle M^2_\Delta \rangle_{\rm osc} \sim \langle M^2_{R} \rangle_{\rm osc}$ is reached before the tachyonic windows become too narrow to support coherent particle production. After the tachyonic growth stops, the strongly non-linear system still undergoes a transient period of resonant particle production driven by the two-point function $\delta \Delta_{\rm F}$ itself, during which $M_{\rm eff}^2$ remains positive. The resonant nature of the particle production can be seen in figure~\cref{fig:rho0k},
which will be discussed further below.  At the onset of the resonance, $M_{\rm eff}^2$ receives roughly equal contributions from the fluctuation term $M^2_\Delta = 3\lambda_{\rm R} \delta \Delta_{\rm F}$ and from the curvature term $M^2_{R} =a^2\bigl(\bar{\xi}_{\mathrm{R}}-1/6\bigr) R$, but as the latter redshifts as $a^{-1}$, it eventually becomes smaller than the fluctuation term. 
The resonance turns off after the effective mass becomes fully dominated by $M^2_\Delta$, and $\delta \Delta_{\rm F}$ on average settles to a constant value.
For $\lambda_{\rm R} = 10^{-4}$ and $10^{-1}$, we find that $\delta \Delta_{\rm F}$ at the end of the tachyonic stage agrees relatively well with the adiabatic expansion results of~\cite{Fairbairn:2018bsw}. However, the subsequent strongly non-linear resonant stage is not at all captured in the treatment of~\cite{Fairbairn:2018bsw} and, as seen in figures~\cref{fig:variance1} and~\cref{fig:meffcomponents}, this stage gives the dominant contribution to $\delta \Delta_{\rm F}$ for $\lambda_{\rm R} = 10^{-4}$ and $10^{-1}$. 
%
%
%
%

The momentum space structure of $\delta\rho_{0\bm k}$ 
is shown in figure~\cref{fig:rho0k}.~For all three coupling values $\lambda_{\rm R} \in \{10^{-7},10^{-4},10^{-1}\}$, the leftmost continuous vertical structures, extending from $|{\bm k}|=0$ to a finite cutoff set by the effective mass (and of the order of the Hubble scale), are states populated by the tachyonic instability. 

For $\lambda_{\mathrm{R}} = 10^{-7}$ the ultraviolet region develops, around $a/a_0\simeq 3$, discrete bands which reach to higher ${|\bm k|}$-modes than the initial structures, while the evolution is still dominated by $M^2_{R}$ (see figure~\cref{fig:meffcomponents}). These bands appear to signal a resonant particle production sourced by the $\xi R\chi^2$-term, which can coexist with the tachyonic production~\cite{Dufaux:2006ee,Bassett:1997az,Cembranos:2019qlm}.~We note that $\delta\rho_{0{\bm k}}$ continues to be strongly dominated by the lowest band but its peak shifts from $|{\bm k}|\approx 0$ towards the middle of the band.

For $\lambda_{\mathrm{R}} = 10^{-4}$ and $10^{-1}$ the momentum space evolution looks quantitatively similar as above until the moment when the effective mass gets dominated by the two-point function, $\langle M^2_\Delta \rangle_{\rm osc} > \langle M^2_{R} \rangle_{\rm osc}$, and $\delta \Delta_{\rm F}$ starts to grow rapidly (see figures~\cref{fig:meffcomponents} and~\cref{fig:variance1}). At this point, pronounced band structures emerge in figure~\cref{fig:rho0k}, which we interpret to signal the onset of resonant particle production driven by $\delta \Delta_{\rm F}$ itself. The resonance bands carry significant power and extend considerably above the ${|\bm k|}$-region populated during the $M^2_{R}$-dominated stage. Furthermore, it can be seen that the moment at which the resonant growth effectively stops in figure~\cref{fig:variance1} corresponds to a further splitting and narrowing down of the resonance bands in figure~\cref{fig:rho0k}. After this band splitting the resonant particle production loses efficiency and the average value of $\delta\Delta_{\rm F}$ becomes essentially a constant.
%
%
\begin{figure}[t]
\centering
\includegraphics[width=0.95\linewidth]{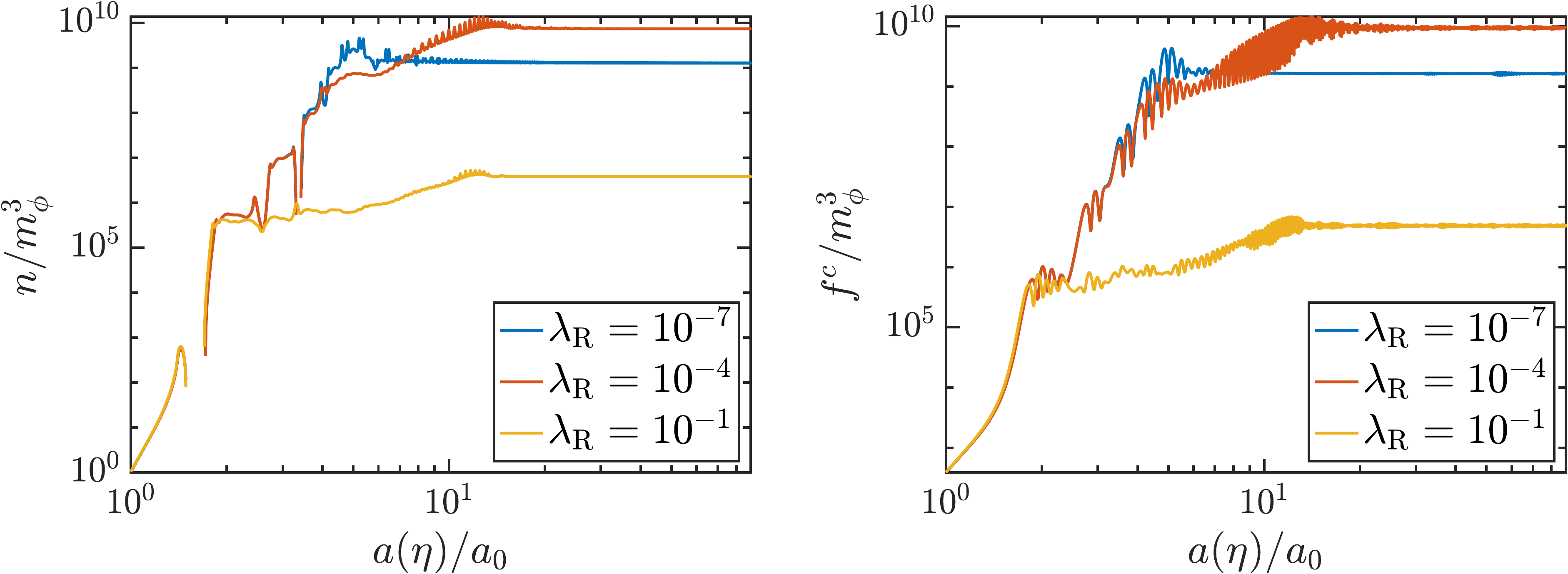}
\caption{The integrated comoving particle number density $n$ (left panel) and the integrated absolute value of the coherence functions $f^c$ (right panel) for $\lambda_{\mathrm{R}} \in \{10^{-7}, 10^{-4}, 10^{-1}\}$ with $\bar\xi_{\mathrm{R}} = 50$ and $\Gamma = 0$.}
\label{fig:nandfc}
\end{figure}
%
%

The evolution of the comoving particle number density $n$ and the coherence function $f^{c}$ are shown in figure~\cref{fig:nandfc}. For $\lambda_{\mathrm{R}} = 10^{-7}$ both $n$ and $f^{c}$ settle to constant values after the end of the tachyonic growth. Comparing with \cite{Fairbairn:2018bsw}, we find an order of magnitude smaller final number density for $\lambda_{\rm R} = 10^{-7}$, the reason being the same as for the difference in $\delta\Delta_{\rm F}$ discussed above. On the other hand, for $\lambda_{\mathrm{R}} \in \{10^{-4}, 10^{-1}\}$ the tachyonic stage is followed by a transient resonance, during which $n$ and $f^{c}$ grow further, and the resonant contribution actually dominates their final values. In these cases our results for the net particle number density exceed the corresponding results of~\cite{Fairbairn:2018bsw} by an order of magnitude. Note that the particle production is necessarily associated with a growing  coherence function~\cite{Fidler:2011yq}. The fact that coherence remains constant after particle production ends shows that the final state is highly squeezed. This is a special feature of our non-interacting system. In an interacting system the coherence function would eventually tend to zero, reducing the quantum system to a non-coherent statistical state, even if the interactions were conserving the particle number. Such behaviour was indeed observed and studied in detailed in a toy model in~\cite{Kainulainen:2021eki}.
%
%
\begin{figure}[h!]
\centering
\includegraphics[trim={2.5cm 0 4cm 0},clip,width=0.95\linewidth]{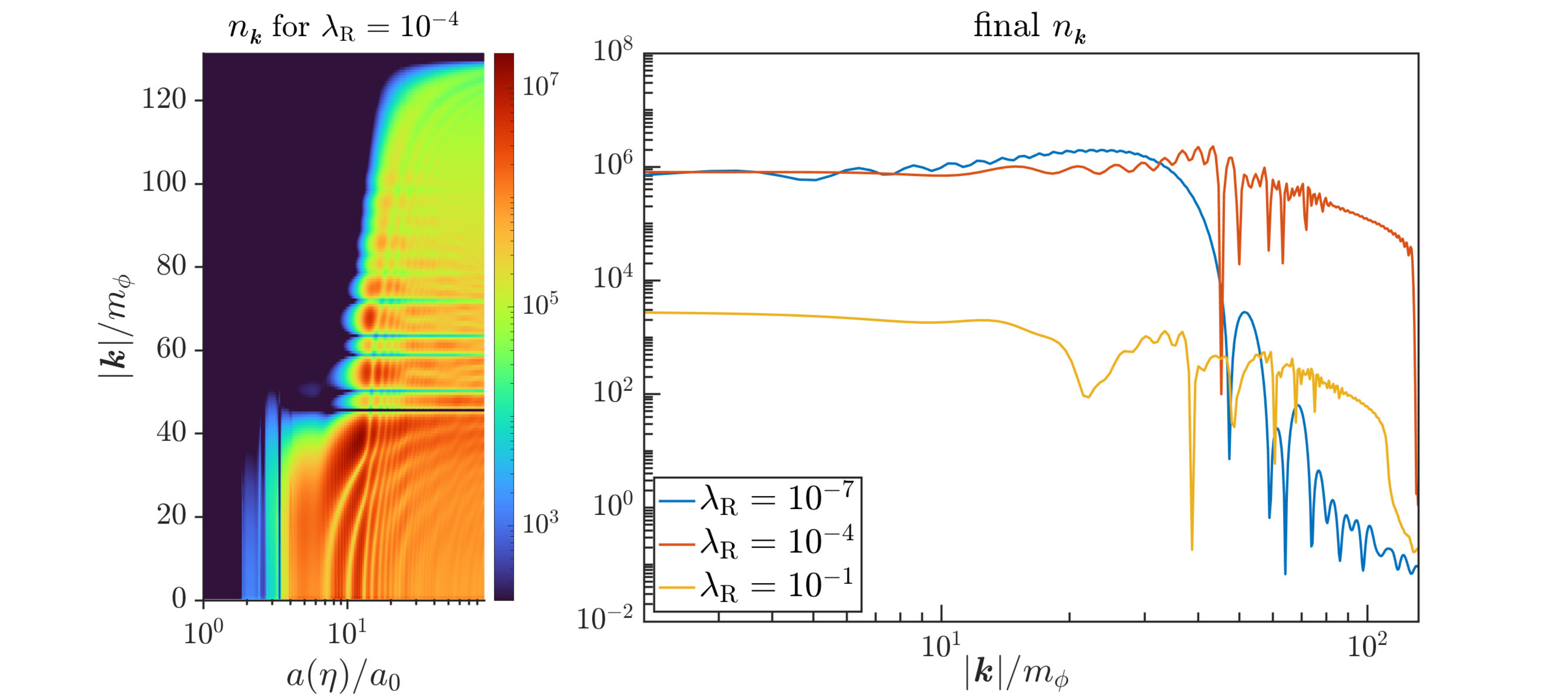}
\caption{A contour plot of the comoving particle number density $n_{\bm k}$ for $\lambda_{\mathrm{R}} = 10^{-4}$ (left panel), and the final comoving particle number density $n_{\bm k}(\eta_{\mathrm{end}})$ as a function of momentum for $\lambda_{\mathrm{R}} \in \{10^{-7}, 10^{-4}, 10^{-1}\}$ (right panel). Both plots have $\bar\xi_{\mathrm{R}} = 50$ and $\Gamma = 0$.}
\label{fig:nfinal}
\end{figure}
%
%
%
%

In figure~\cref{fig:nfinal} we show the comoving particle number density per momentum $n_{\bm k}$. The right panel shows the final spectrum $n_{\bm k}$ at the final time of our numerical simulation for all couplings considered: $\lambda_{\mathrm{R}} \in \{10^{-7}, 10^{-4}, 10^{-1}\}$.  The left panel shows the full time evolution of $n_{\bm k}$ for the coupling $\lambda_{\mathrm{R}} = 10^{-4}$.    
Apart from the oscillatory features, the structure of $n_{\bm k}$ is qualitatively in agreement with the results of~\cite{Fairbairn:2018bsw}, which, we recall, are obtained using a semianalytical adiabatic expansion approximation for the tachyonic particle production~\cite{Dufaux:2006ee} and neglecting all resonant particle production (see also~\cite{Cembranos:2019qlm} for an analysis of resonant production through the $\xi R \chi^2$-term in the absence of self-couplings). The oscillatory features in $n_{\bm k}$ seen in our results arise from the transient resonance after the first tachyonic stage. As seen in the left panel of figure~\cref{fig:nfinal}, $n_{\bm k}$ displays strong peaks coinciding with the onset of the resonance, located at the resonance bands and with the peak heights varying from band to band. Interestingly, the peaks begin to flatten out while the resonance is still ongoing. This effect is caused by non-linear processes mediated by the self-coupling which, combined with the redshifting, can efficiently redistribute the momenta.

%
%
\begin{figure}
\centering
\includegraphics[width=0.90\linewidth]{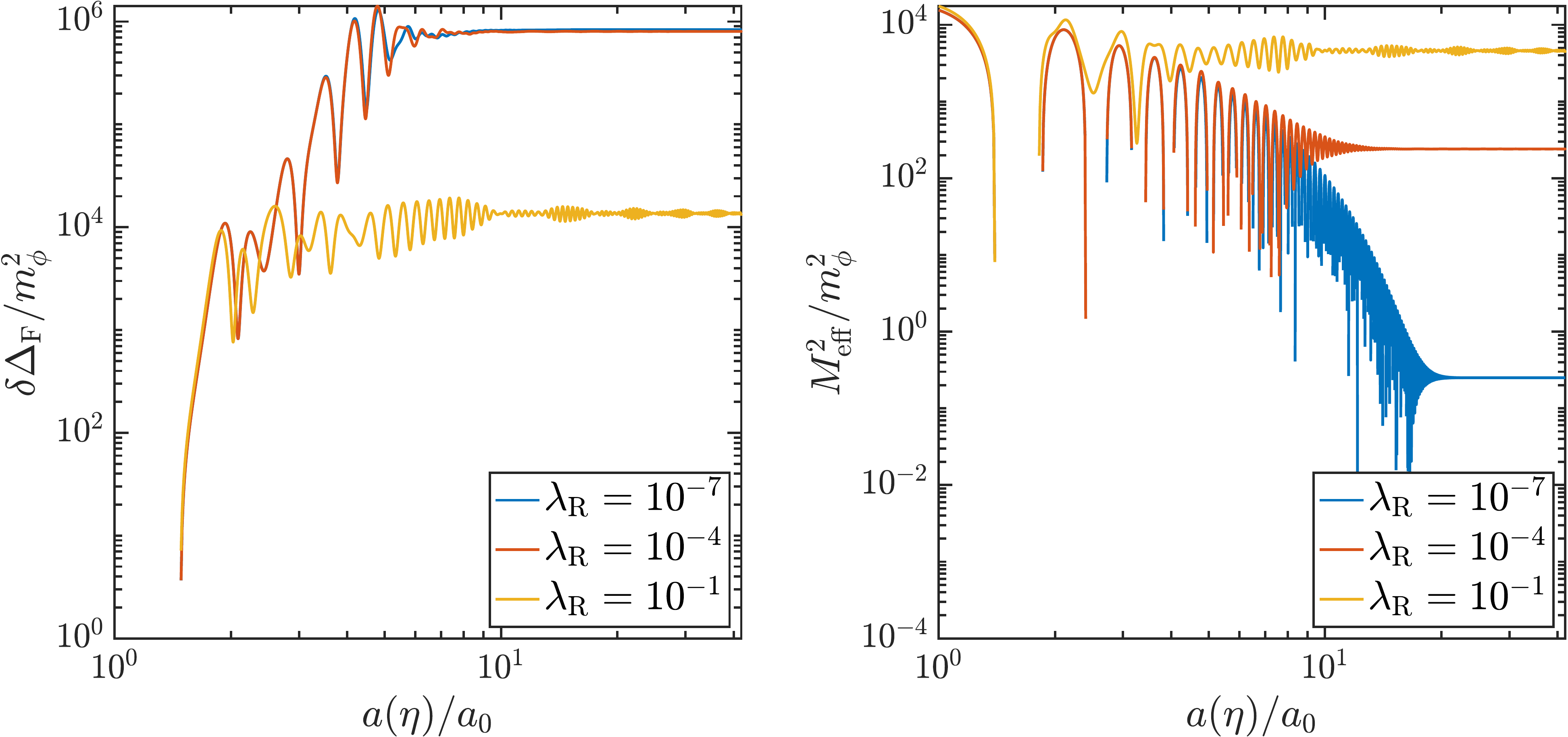} 
\caption{The two-point function $\delta\Delta_{\mathrm{F}}$ (left panel) and the effective mass function \(M^2_{\mathrm{eff}}\) (right panel). The results are shown for $\lambda_{\mathrm{R}} \in \{10^{-7}, 10^{-4}, 10^{-1}\}$, $\bar\xi_{\mathrm{R}} = 50$ and $\Gamma \simeq 0.1H_0$.}
\label{fig:varianceII}
\end{figure}
%
%

\paragraph{Case II: {$\bm{\bar{\xi}_{\mathrm{R}} = 50, \Gamma \simeq 0.1 H_0}$.}} 

\begin{figure}
\centering
\includegraphics[width=0.90\linewidth]{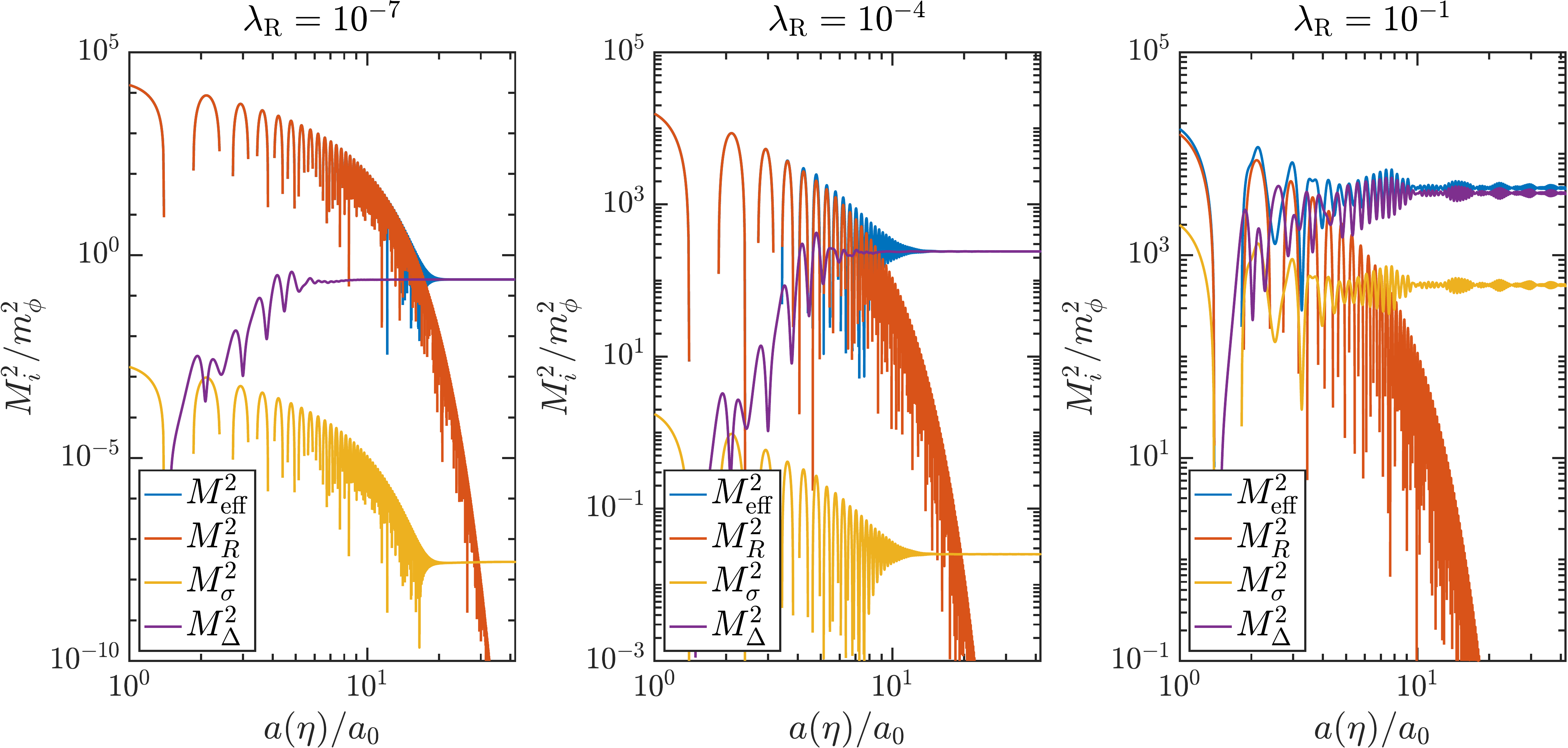}
\caption{The effective mass function $M^2_{\mathrm{eff}}$ (blue) and its component functions $M^2_{R}$ (red), $M^2_{\Delta}$ (violet) and $M^2_{\sigma}$ (yellow), defined in equations~\cref{eq:Meff_components}, for $\lambda_{\mathrm{R}} \in \{10^{-7}, 10^{-4}, 10^{-1}\}$ in the case $\bar\xi_{\mathrm{R}} = 50$ and $\Gamma \simeq 0.1H_0$.}
\label{fig:meffcomponentsII}
\end{figure}

For comparison, we also present results for the case with $\bar{\xi}_{\mathrm{R}} = 50$ and a non-zero inflaton decay rate $\Gamma \simeq 0.1 H(\eta_0)\equiv 0.1 H_0$. As explained in section~\cref{sec:model}, inflaton decays into radiation, as a result of which the universe evolves from effective matter domination to radiation domination where $R = 0$. The evolution of $\delta\Delta_{\rm F}$ and $M_{\rm eff}^2$, and the components of $M_{\rm eff}^2$ defined in equations~\cref{eq:Meff_components}, are shown in  figures~\cref{fig:varianceII} and \cref{fig:meffcomponentsII} for this case. As is seen in figure~\cref{fig:meffcomponentsII}, the initial scaling $\langle R\rangle_{\rm osc} \propto a^{-3}$ is now followed by an exponential decay of $\langle R\rangle_{\rm osc}$ once the inflaton decay becomes efficient. This decreases the efficiency of tachyonic particle production compared to case I.

%
%

The evolution of $\delta\Delta_{\mathrm F}$ seen in figure~\cref{fig:varianceII} is now almost identical for the couplings $\lambda_{\rm R} = 10^{-7}$ and $10^{-4}$. This is due to the fast decrease of $R$ resulting from the inflaton decay, which ends the tachyonic growth before the two-point function starts to backreact into the dynamics also for  $\lambda_{\mathrm R} = 10^{-4}$. This can also be seen from figure~\cref{fig:meffcomponentsII}, which shows that in both these cases $\delta\Delta_{\rm F}$ stops growing before the two-point function backreacts into the dynamics. The evolution of $\delta\Delta_{\rm F}$ for $\lambda_{\rm R} = 10^{-7}$ is qualitatively similar to case I, but the final value of $\delta\Delta_{\rm F}$ is about two orders of magnitude smaller. For $\lambda_{\rm R} = 10^{-4}$, the evolution of $\delta\Delta_{\rm F}$ substantially differs from case I as the resonant stage that dominated the final value of $\delta\Delta_{\rm F}$ in case I is absent in case II. For the largest coupling $\lambda_{\mathrm{R}}= 10^{-1}$ the difference compared to case I is smallest as the tachyonic growth in this case still terminates via the backreaction when $\langle M^2_\Delta \rangle_{\rm osc} \sim \langle M^2_{R}\rangle$, and this happens before the exponential decrease of $R$ sets in. In this case, the tachyonic stage is followed by resonant amplification of $\delta \Delta_{\rm F}$ driven by $\delta \Delta_{\rm F}$ itself, but the resonance is somewhat less efficient than in case I, leading to a factor of two smaller final value for $\delta \Delta_{\rm F}$. 

\begin{figure}[t]
\centering
\includegraphics[trim={3.1cm 0 3.1cm 0},clip,width=1\linewidth]{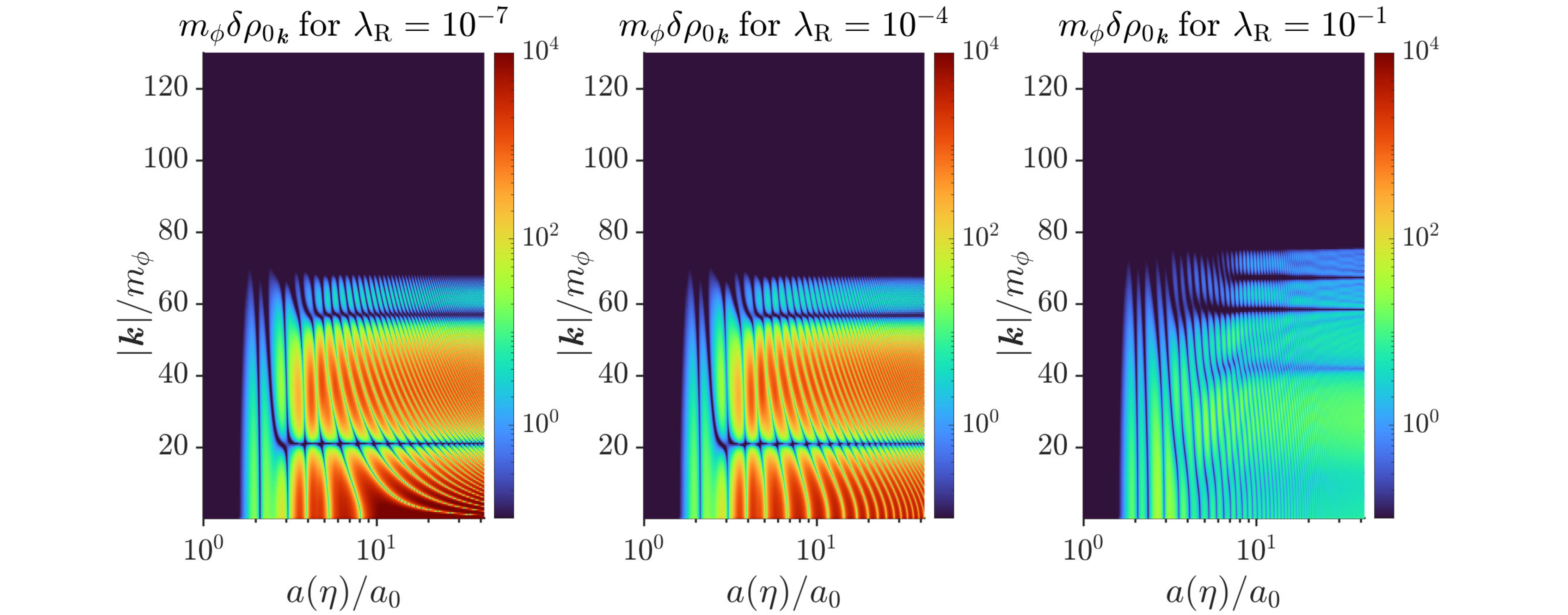}
\caption{The zeroth moment $\delta\rho_{0 \bm k}$ of the two point function for $\lambda_{\mathrm{R}} \in \{10^{-7}, 10^{-4}, 10^{-1}\}$ with $\bar\xi_{\mathrm{R}} = 50$ and $\Gamma \simeq 0.1H_0$.}
\label{fig:rho0kII}
\end{figure}

Finally, the momentum structure of $\delta\rho_{0\bm k}$ is shown in figure~\cref{fig:rho0kII}.~For $\lambda_{\mathrm{R}}= 10^{-7}$ the result looks qualitatively similar to case I but the band structures generated during the $M^2_{R}$-dominated epoch are more pronounced in case II. In particular, in case II the tachyonic region splits into two discrete bands at $a/a_0 \simeq 3$. The results for $\lambda_{\mathrm{R}}= 10^{-4}$ look almost identical to those for $\lambda_{\mathrm{R}}= 10^{-7}$, and the $\delta \Delta_{\rm F}$-driven resonance that dominated the final $\delta\rho_{0\bm k}$ in case I is now completely absent. For $\lambda_{\mathrm{R}}= 10^{-1}$ the structure looks qualitatively similar to case I but it can be seen that the $\delta \Delta_{\rm F}$-driven resonance is less efficient and does not extend to as high momenta as in case II.
%
%

All in all, the results of cases I and II manifest the presence of complicated non-linear dynamics after the initial tachyonic particle production which can substantially affect the final value of $\delta\Delta_{\rm F}$. In particular, our results indicate that when the two-point function grows large enough to backreact into the dynamics, the tachyonic instability is followed by resonant particle production driven by the two-point function itself. In all cases studied here, we find that if the resonance takes place it also gives a dominant contribution to the final value of $\delta\Delta_{\rm F}$. However, the amount by which $\delta\Delta_{\rm F}$ grows during the resonance after the tachyonic stage appears to depend quite sensitively on the non-linear evolution of the two-point function coupled to $R$.

%
\section{Conclusions}
\label{sec:conclusions}
%

We have studied particle production at the end of inflation with a non-minimally coupled spectator scalar field that contributes to dark matter. We first introduced consistently renormalized coupled equations for the one- and two-point functions of the spectator field in the Hartree approximation using 2PI-methods. These equations correctly account for the backreaction of the out-of-equilibrium quantum modes created by the spinodal instability triggered by the oscillating Ricci scalar as well as for the subsequent parametric resonances. This model was studied earlier in~\cite{Fairbairn:2018bsw} with an adiabatic treatment of the spinodal effects. Our results show that the interplay between the backreacting two-point function and the oscillating curvature sector lead to highly non-trivial dynamics which can have a significant effect on the net particle number density.

We solved numerically the coupled equations for the one- and two-point functions of the spectator field (the latter expressed as moment equations in the Wigner representation) together with the dynamical evolution of the inflaton sector for different values of the spectator field self-coupling $\lambda_{\mathrm{R}}$ and for the minimal coupling $\bar{\xi}_{\mathrm{R}} = 50$. We studied first the case of a non-interacting inflaton field and found that for a small coupling $\lambda_{\mathrm{R}} = 10^{-7}$ the generated particle number density is an order of magnitude smaller than that found in~\cite{Fairbairn:2018bsw}, whereas for $\lambda_{\mathrm{R}} = 10^{-4}$ and $10^{-1}$ it becomes and order of magnitude larger. For $\lambda_{\mathrm{R}} = 10^{-7}$ this is due to the tachyonic particle production shutting off already before the competing mass contributions from the curvature and the two-point function become comparable, while for the larger couplings the difference is due to efficient resonant particle production occurring after the tachyonic stage. In particular the resonant production, which actually dominates the contribution to the particle number density for larger couplings, is completely absent in the adiabatic approach of~\cite{Fairbairn:2018bsw}.

We also included a coupling between the inflaton and a radiation component to study the evolution under the transition from effective matter domination to radiation domination with $R = 0$. We found that the exponential decay of $R$ induced by the radiation coupling renders both the spinodal and the resonant particle production processes much less efficient compared to the case with a non-interacting inflaton. For the tachyonic processes this is easy to understand as the oscillating curvature term, which is responsible for the tachyonic bursts in the particle number density, is rapidly driven to zero. Our results suggest the presence of an $R$-assisted resonance enhancement, where the resonant particle production driven by the two-point function is boosted by the decaying $\xi R \chi^2$-term after the tachyonic stage has come to an end. This is a highly non-linear phenomenon which, when present, appears to dominate the net particle production. It cannot be properly captured without a full treatment of the backreaction effects.

The final momentum distribution of the dark relics generated by the non-perturbative processes is highly non-thermal. This could lead to characteristic and potentially observable imprints in the structure formation, as pointed out in~\cite{Fairbairn:2018bsw}.~The evolution of the relic distribution after the epoch of reheating depends on dark sector interactions, possibly including new types not considered here. Although this would be an interesting problem in itself, we do not investigate it further here.   

It would obviously be interesting to extend our setup to the case of a spectator field coupled to other matter fields. This could be done rather easily by combining the current results with the quantum transport formalism for interacting fermions introduced in~\cite{Jukkala:2019slc}. Also, it would be interesting to extend our classical treatment of the inflaton to quantum level. It would then be particularly interesting to study the gravitational wave production during the reheating stage in the most general computational framework described above.

%
\section*{Acknowledgements}
\label{sec:ack}
%

 This work was supported by the Academy of Finland grant 318319 and by computer capacity 
 from the Finnish Grid and Cloud Infrastructure (persistent identifier urn:nbn:fi:research-infras-2016072533). 
 OK was in addition supported by a grant from the Magnus Ehrnrooth Foundation. We wish to thank Anna Tokareva for many useful discussions and comments on the manuscript.

%
\bibliography{maindesc.bib}
%

%
\end{document}